\documentstyle[aps,preprint,epsf]{revtex}
\tightenlines 
\begin{document}
\title{Far-infrared study of the insulator-metal transition 
in $\theta$-(BEDT-TTF)$_{\bf {2}}$RbZn(SCN)$_{\bf {4}}$}
\author{N. L. Wang}
\address{Institute of Physics and Center for Condensed Matter Physics, 
Chinese Academy of Sciences, P.O.Box 2711, Beijing 100080, P. R. China}
\author{H. Mori and S. Tanaka}
\address{Superconductivity Research Laboratory, ISTEC,
 Shinonome 1-10-13, Tokyo 135-0062, Japan} 
\author{J. Dong and B. P. Clayman}
\address{Department of Physics, Simon Fraser University, Burnaby, 
British Columbia, V5A 1S6, Canada}
\date{Received Feb. 1, 2001, revised Apr 26, 2001}
\maketitle
\begin{abstract} 
The temperature-dependent infrared properties of $\theta$-(BEDT-TTF)$_2$RbZn(SCN)$_4$ 
were investigated. In the temperatures above the metal-insulator (M-I) transition, 
the optical conductivity remains finite in the low energy region, but shows no 
conventional Drude feature typical of metals. Below the M-I transition temperature, the 
low-energy spectral weight is significantly reduced. An opening of optical gap at 
about 300 cm$^{-1}$ is observed clearly in the polarization parallel to the 
donor-stacking direction. Analysis of both the electronic and vibronic spectra 
suggests a redistribution of charges in ET molecules as temperature decreases 
across the M-I transition, providing strong support for the charge-ordering state 
at low temperatures.
\end{abstract} 


\epsfclipon
Among various organic compounds containing electron-donor BEDT-TTF (abbreviated 
as ET) molecules, the $\theta$-type family $\theta$-(ET)$_{2}$X, 
where X is an anion, recently attracted much attention because they show 
various unusual physical phenomenen. The compounds have a two-dimensional 
(2D) layer structure consisting of ET donor sheets and insulating anion 
X sheets (in the ac plane). Within the donor plane, the donor stacks regularly 
along the c direction. A dihedral angle $\theta$ is formed between ET 
molecules in neighboring donor columns, which is different for each compound. 
\cite{Mori1} For monovalent X$^-$, half a hole is 
introduced into one ET$^{1/2+}$ molecule, which makes a quarter-filled band. 
On the basis of the band picture, one would expect a metallic behavior in 
those compounds. However, most of the $\theta$-type compounds undergo very sharp 
metal-insulator (M-I) transitions with decreasing temperature.  For 
$\theta$-(BEDT-TTF)$_2$MM'(SCN)$_4$ [M=Tl, Rb, M'=Zn, Co], 
the resisitivities jump several orders at the transition temperatures. \cite{Mori1} 
It is of great interest to understand the origin of the M-I 
transition in these compounds.

It is well-known that a strong electron correlation can lead to a 
metal-insulator transition. For a half-filled system, the strong 
on-site Coulomb interaction would prevent two electrons from occupying 
the same site, which leads to the so-called Mott metal-insulator 
transition. The antiferromagnetic (AF) insulating state in the 
$\kappa$-type ET salts can be recognized as the Mott insulating state, 
since the ET molecules are strongly dimerized which effectively makes the 
system have one hole per dimer. \cite{Kino,McKenzie} For the case 
of the quarter-filled band, the mechanism of the M-I transition 
is different. Fukuyama et al. \cite{Fukuyama}
proposed that the mutual Coulomb interactions of not only on-site 
but also between different sites play important role in this situation, 
leading to a state where electrons are localized every other sites 
as in the Wigner crystal. This is a kind of charge ordering (CO) 
state, which provides a route to the correlation-induced insulator 
transition. Recent NMR measurements \cite{Miyagawa,Chiba} 
on $\theta$-(BEDT-TTF)$_2$RbZn(SCN)$_4$ (abbreviated as $\theta$-RbZn)
indicate the existence of two kinds of donor molecules with different 
amount of charges below the metal-insulator transition temperature 
(T$_{MI}$) at 190 K, implying a CO state in the low-temperature 
insulating state. 

Optical spectroscopy is an important technique for probing the 
electronic state of a material. Tajima et al. performed the first 
optical reflectance measurement of $\theta$-type ET salts. 
By comparing the extracted optical conductivity spectra with 
those calculated by means of the mean-field approximation, they concluded 
that the $\theta$-RbZn salt has horizontal charge ordering below 
T$_{MI}$. \cite{Tajima} However, the reflectance measurement 
did not extend to the far-infrared region. In this case, the 
low-frequency extrapolation in the Kramers-Kronig transformation 
may affect the resultant spectra of conductivity. In addition, the 
vibronic features across the transition were not discussed. In this 
work we present the far-infrared reflectance measurement on 
$\theta$-(BEDT-TTF)$_2$RbZn(SCN)$_4$. We find that the spectral change 
with decreasing temperature is different for the polarizations parallel 
and perpendicular to the donor stacking direction. An opening of gap 
about 300 cm$^{-1}$ below T$_{MI}$ is clearly observed in the 
polarization parallel to the c-axis, but a smaller gap magnitude is 
seen for $\bf E$ parallel to the a-axis. We shall see that 
both the electronic and vibronic spectra suggest redistribution of charges 
in ET molecules below the M-I transition temperature, 
providing further support for the CO state at low temperatures. 

Salts of $\theta$-(BEDT-TTF)$_2$RbZn(SCN)$_4$ were prepared by an 
electrochemical method. \cite{Mori1} Optical reflectivity spectra were 
measured from 50 cm$^{-1}$ to 9500 cm$^{-1}$ on a Bruker 113v spectrometer 
using an {\em in-situ} overcoating technique. \cite{Homes} In order to 
avoid a supercooled state of high-T metallic phase, the 
cooling rate between 200 K and 180 K is very slow---about 0.1 K/min. 
The optical conductivity was calculated via a Kramers-Kronig analysis of the 
reflectivity. Appropriate extrapolations---the Hagen-Rubens relation for 
measurement at high-T and constant value for measurement at low-T---were used 
below 50 cm$^{-1}$ for this analysis. In fact, the conductivity 
in the measured frequency range was found to be insensitive to different 
extrapolations at low $\omega$. For the high-frequency region, we 
extrapolated the measured spectrum at 9500 cm$^{-1}$ as constant to 
150000 cm$^{-1}$, above which a $\omega^{-4}$ relation was employed.

Fig. 1 shows the room temperature reflectance and conductivity spectra 
of $\theta$-(BEDT-TTF)$_2$RbZn(SCN)$_4$ in polarizations $\bf E \| c$ and 
$\bf E \| a$ . The spectra exhibit several remarkable features. First, the 
reflectance is much less than unity at low frequencies in both 
polarizations. The calculated conductivity spectra exhibit a broad 
electronic band centered around 2000 cm$^{-1}$ for both polarizations. 
Such feature was observed in other organic salts, and was suggested 
by us to originate from the formation of small polarons. \cite{Wang}
The spectra are far from the conventional Drude response typical of 
metals. Nevertheless, the optical conductivity remains finite in the 
limit of $\omega \rightarrow$ 0. The observation indicates that the 
material at high temperature, though not a good metal, is not a gaped 
insulator as well. Second, there exists considerable spectral 
anisotropy over a wide frequency range. The reflectivity in the {\em 
a} direction is significantly larger than that in the donor stacking 
direction (c-axis). Consequently, the conductivity has a higher 
spectral weight in the polarization of $\bf E \| a$. This is consistent with 
the theoretical calculation that the transverse transfer integral 
(t$_p$) is larger than that along the stacking direction (t$_c$). 
\cite{Mori1} Third, the vibrational response displays distinctive 
electron-phonon coupling characteristics. Except for a sharp phonon 
appearing at ~2100 cm$^{-1}$ which is attributed to the C-N 
stretching mode of the anion, the spectra exhibit several vibrational 
modes superimposed in electronic spectra near ~1350 cm$^{-1}$, 850 
cm$^{-1}$, and 470 cm$^{-1}$ . They have typical Fano line shape 
caused by the strong electron-molecule-vibration (EMV) coupling: an 
antiresonance or dip preceded by a peak on the low-frequency side in 
the infrared conductivity spectrum. The vibronic structure is a 
characteristic feature in organic conductors. \cite{Jacobsen}

The temperature-dependent reflectivity and conductivity spectra for 
the polarization parallel to the donor stacking direction ($\bf E \| c$) is 
shown in fig. 2. The spectra above T$_{MI}$ do not 
show much T-dependence. However, dramatic change occurs below 190 K. 
The reflectivity is significantly reduced in the far-infrared region, 
but enhanced somewhat at frequencies higher than 1500 cm$^{-1}$. As 
a result, a redistribution of spectral weight in $\sigma_1(\omega)$ 
is observed as the temperature decreases across 190 K. The missing 
spectral weight below 2000 cm$^{-1}$ is transferred to the frequencies 
higher than 2000 cm$^{-1}$.  The calculated optical conductivity has almost 
no spectral weight at very low frequencies, suggesting an opening of 
gap below T$_{MI}$. The gap magnitude in the donor stacking direction is estimated to be about 300 cm$^{-1}$ by a linear extrapolation of 
the conductivity onset as indicated 
by a solid line in the figure. Another remarkable observation is that, 
accompanying the redistribution of the electronic excitations, most of 
antiresonances of EMV modes become the ordinary (Lorentzian) vibrational 
peaks, whose maxima correspond to the minima of the antiresonances. In 
addition, a new dip appears in the conductivity spectra at around 
2800 cm$^{-1}$.

Fig. 3 shows the spectra for the polarization $\bf E \| a$ at several 
temperatures. The spectra evolve with temperature in a similar way as 
that of $\bf E \| c$. However, the spectra weight for $\bf E \| a$ 
is reduced in a much broader frequency range in the insulating state. 
Moreover, the conductivity starts to increase at much lower frequency 
in comparison with the polarization of $\bf E \| c$. This 
suggests that the gap magnitude is much smaller in this polarization.
We noticed that our conductivity spectra at low temperature are different 
from the earlier data by Tajima et al. \cite{Tajima} In that work, except 
for the sharp phonon peaks, there is almost no spectral weight below 2000 cm$^{-1}$. Since the 
far-infrared reflectance was not measured in their experiment, the missing 
spectral weight must be due to the low-$\omega$ extrapolation in 
their Kramers-Kronig calculation. 

The observation of an optical gap below T$_{MI}$ is a notable result 
in this work. We emphasize that the differences in the two polarizations 
are not artifacts caused by the Kramers-Kronig analysis of the experimental reflectance data, since different low-$\omega$ extrapolations actually do 
not alter the conductivity spectra in the measured frequency region. 
As we shall see that the anisotropy of optical responses correlates well 
with the different parameters of electronic correlations in the two directions.  

As we mentioned in the introduction, 
recent experimental and theoretical works established a charge-ordering 
transition in this $\theta$-type compound. NMR measurement indicated that the 
ET donors, which are equivalent at high T with a charge amount of 0.5, become
inequivalent at low T. \cite{Miyagawa,Chiba} A disproportionation 
of charges like D$^{0.5+\delta}$D$^{0.5-\delta}$D$^{0.5+\delta}$D$^{0.5-\delta}$... 
occurs at T$_{MI}$. Therefore, there exist two kinds of ET molecules with different amount of charges in the low temperature, which implies a charge ordering below T$_{MI}$. The charge ratio estimated from those NMR measurements is 
about 1:(4 $\sim$ 6), \cite{Miyagawa,Chiba} which gives $\delta$=0.30 $\sim$ 
0.36. On the other hand, X-ray diffraction 
measurement demonstrated that the M-I transition was associated with the lattice 
modulation with a period doubling along the c-axis, \cite{Mori2} {\em i.e.} the ET donors form dimers in the c-axis. These 
experiments suggest that the CO is coupled with the lattice modulation. 
Charge ordering formation was also suggested from theoretical studies, where 
the intermolecular Coulomb repulsion V was suggested to play 
a crucial role. \cite{Fukuyama,Mori3,McKenzie2} This is because the Hubbard 
model only taking account of the on-site Coulomb repulsion energy 
was believed to be always metallic at quarter filling. Mori calculated the 
intersite Coulomb repulsion V as a function of the dihedral angle of the 
ET molecules and found that the intersite Coulomb repulsion V$_c$ along 
the stacking direction is larger than that of V$_p$ along the transverse 
direction in most of the actual $\theta$-type salts. The CO takes 
the direction of the smaller intersite Coulomb repulsions in the $\theta$-type 
salts.\cite{Mori3} Taking account of both the experimental and theoretical 
studies on the $\theta$-type ET salts, the electronic states of ET donors 
above and below T$_{MI}$ in the conducting layer can be summarized in 
fig. 4 (we assume $\delta$=0.3 here). \cite{Mori4} 

We argue that our anisotropic optical responses are consistent 
with the horizontal charge ordering picture in fig. 4. Upon cooling down the sample, the dihedral angle increases because of the thermal contraction.\cite{Mori3} The intersite Coulomb repulsion V$_c$ becomes 
effective when the lattice modulation along c-axis takes place below 
190 K, as a result, the horizontal charge ordering forms. Since the Coulomb repulsion energy V$_c$ is larger than V$_p$, meanwhile the charge transfer 
integral t$_c$ is smaller 
than t$_p$, the electrons have to overcome larger barrier potentials in the c-direction than in the a-direction for transport. This is in agreement with the experiment that a larger gap is observed in the c-direction. The gap size 
is found to be scarcely reduced with increase of temperature up to T$_{MI}$, 
which is very similar to other strongly correlated systems with CO transitions, 
for example, the Fe$_3$O$_4$ compound.\cite{Park} As can be seen from 
fig. 4, a charge density wave (CDW) is actually formed along the c-direction. 
Therefore, the development of the optical gap below T$_{MI}$ in the present 
CO state may have the similar origin as in the CDW gap. However further 
theoretical work is needed to elucidate the mechanism of the gap formation in 
the CO state of the material. 

We will now turn to the discussion of the vibrational features. It is 
well established that most of the observed phonon features are due to the 
totally symmetric A$_g$ modes of intramolecular vibrations which coupled 
strongly to the charge densities distributed on the donor 
molecules. \cite{Eldridge,Dong,Sugano} Thus, the 
dramatic change of the vibronic structure in the course of M-I transition also 
reflects significant redistribution of charges on ET molecules. This is also 
consistent with the CO picture.

The assignments of the vibrational modes of ET-based salts were thoroughly 
investigated in previous studies. \cite{Eldridge,Dong,Sugano} In the compound 
under study, except for the sharp C-N stretching mode from anion at ~2100 cm$^{-1}$, 
major vibrational features occur near ~1350 cm$^{-1}$ (very broad), 850 cm$^{-1}$, 
and 470 cm$^{-1}$. Comparing with those works, we can assign the strong and 
broad structure near 1350 cm$^{-1}$ to the central C=C stretching mode $\nu_3(a_g)$, 
the feature near ~470 cm$^{-1}$ to the C-S stretching modes $\nu_9(a_g)$ and $\nu_{10}(a_g)$. 
However, the feature at ~850 cm$^{-1}$ should be assigned to the $\nu_{60}(B_{3g})$ 
normal mode involving the motion of four of the inner-ring carbon atoms, 
rather than to $\nu_7(a_g)$ mode. \cite{Eldridge} As noted above, those vibronic modes 
of ET donors change from the antiresonances at high T to the ordinary 
(Lorentzian) peaks at low T. Moreover, from fig. 2 and fig.3, we find that the 
line shape is related to the relative strength of the electronic excitations: 
the antiresonance takes place if the vibrational mode overlaps with a higher 
electronic background; whereas the ordinary peak appears when the electronic 
background becomes lower. We believe that the change of the vibrational structure 
is due to the change of the screening action of mobile charges, caused by 
either more localized electrons or the smaller charge density around the bonds associated 
with those vibrational modes below T$_{MI}$.

Another notable observation is that a new dip appears at frequency near 
2800 cm$^{-1}$ for both polarizations below T$_{MI}$. The feature was also 
observed in previous measurement by Tajima et al., where it was considered to be 
formed by two separated electronic bands. In our opinion, the dip is more 
likely related to the vibronic structure, i.e. an antiresonance of a vibrational mode.  
We assign it to the C-H stretching mode $\nu_1(a_g)$, which was usually very 
weak in the ET-based salts. \cite{Eldridge} Similar feature was also observed in
some other ET-based compounds at the same frequency region. \cite{Dressel,Baker} The
sizable strength of this antiresonance feature suggests a significant EMV
coupling for this mode in comparison with other ET salts.  This could be
due to a higher charge density distribution around the H atoms upon CO
transition at low temperature. Taking account of both the low temperature
emergence of the remarkable C-H stretching vibrational feature and the change 
of other vibrational modes, we also draw the
conclusion that there is a charge redistribution across the metal-insulator
transition. A conceivable scenario is :  a certain amount of charges 
were transferred from around the inner-ring atoms of a ET 
molecule to the place around C-H bonds in the outer-ring of another ET molecule, thus making 
the disproportionation of charges as depicted in fig. 4. But these localized charges 
greatly reduce the low-$\omega$ conductivity spectral weight. Additionally, we find that 
the number of vibrational peaks increases in the low T, which is apparently due to the 
structure modulation occurred at T$_{MI}$.

To conclude, our far-infrared spectroscopy study revealed dramatic change of both 
electronic and vibrational spectra across the metal-insulator transition. The change 
can be well understood from the charge-ordering transition, which is consistent with 
the result of earlier NMR measurements. 
   
We thank T. Timusk, Y. P. Wang and S. P. Feng for valuable discussions. This 
work was supported by research grants from NNSFC (No. 19974049), Simon Fraser 
University, and NEDO.

\begin{figure}[htb]
\vspace{0.1in}
\caption{The frequency dependence of reflectivity and 
conductivity for $\bf E \| c$ and $\bf E \| a$ at room temperature.} 
\label{1}
\end{figure}
\begin{figure}[htb]
\vspace{0.1in}
\caption{The frequency dependence of reflectivity and 
conductivity for $\bf E \| c$ at different temperatures.} 
\label{2}
\end{figure}
\begin{figure}[htb]
\vspace{0.1in}
\caption{The frequency dependence of reflectivity and 
conductivity for $\bf E \| a$ at different temperatures.} 
\label{3}
\end{figure}
\begin{figure}[htb]
\vspace{0.1in}
\caption{A schematic picture of electronic 
states of ET donors above and below T$_{MI}$ in the conducting 
layer.} 
\label{4}
\end{figure}

\end{document}